# From Dumb Wireless Sensors to Smart Networks using Network Coding

Alexandros G. Dimakis, Dragan Petrovic and Kannan Ramchandran
Department of Electrical Engineering and Computer Science
University of California, Berkeley
{adim, dragan, kannanr}@eecs.berkeley.edu

## I. INTRODUCTION

The vision of wireless sensor networks is one of a smart collection of tiny, dumb devices. These motes may be individually cheap, unintelligent, imprecise, and unreliable. Yet they are able to derive strength from numbers, rendering the whole to be strong, reliable and robust. In the face of scaling the network under uncertainty (due to a combination of fading channels, interference and dynamic environments), it may be overreaching to demand precise global network coordination and knowledge. Likewise, given that devices need to get smaller and cheaper, it may be unwise to depend on precisely tuned and expensive components. As node sizes get smaller while the network gets larger, individual capabilities in terms of communication range, memory requirements and computing capabilities cannot grow with network size. Instead there is need for collections of nodes to pool together their meager and unreliable resources, and still form a reliable and robust network.

Our approach is to adopt a distributed and randomized mindset and rely on in-network processing and network coding [3]. Our general abstraction is that nodes should act only locally and independently, and the desired global behavior should arise as a collective property of the network. In this paper we summarize our work and present how these ideas can be applied for communication and storage in sensor networks. Specifically, in Section 2 we summarize our work on how randomized network coding can be used to ensure reliable communication over unreliable, low–power radios. In Section 3, we summarize our work on creating a reliable network memory using decentralized erasure codes in sensor networks.

## II. OVERCOMING UNTUNED RADIOS USING NETWORK CODING

To make sensor network deployments economically and technologically feasible, it is necessary to drastically reduce the cost, size and energy consumption of the nodes available today. Moore's Law still provides for exponential reduction of these metrics over time when it comes to the digital components that comprise the memory, computation and coding of the nodes. However, there is no equivalent trend to Moore's Law that applies to the analog components needed for the radios that enable the nodes to communicate with one another. This work introduces an architecture for the analog radios that can greatly reduce the cost, size (5x reduction) and energy consumption (10x reduction) of the nodes. In fact it is expected that the proposed architecture will allow the energy consumption of the nodes to be so low that they could be fully powered by energy scavenged from the environment [13]. The penalty for using such a radical architecture is that the radios become untuned and it is no longer possible to guarantee that any arbitrary pair of nodes will be able to communicate with each other. Instead, it becomes necessary to rely on the *density* of nodes to make the overall network capable of providing reliable communication.

Narrowband radios have shown to be the architecture of choice for low–power applications [1], [2], as they are low in complexity and consume less power than spread spectrum or other wideband techniques. One fundamental requirement of narrowband radios is that the transmitter's carrier frequency and the receiver's detection frequency must be well matched. This is traditionally accomplished by employing a crystal at both the transmitter and receiver to provide the same low frequency reference. This reference frequency is multiplied via a phase–locked loop (PLL) to generate the carrier wave. However, the off–chip crystal contributes significantly to the cost, size, and power consumption of such transceivers. Therefore, great savings in all three of these areas could be obtained by eliminating the off–chip crystal and PLL. We propose to eliminate the quartz crystal and replace it by an on–chip (LC) circuit to provide the frequency reference. This makes it possible to economically produce millions of nodes and densely deploy them. The proposed architecture allows a sensor node to be developed entirely out of thin–film technologies on a single chip. However, the drawback of such architectures is that the variations in the manufacturing process are large, resulting in untuned radios. Therefore, two narrowband radios produced by such a process are not likely to be able to communicate with each other. To address this problem, we propose to make use of the density of nodes to ensure reliable communication within the network.

### A. Main results

We consider using the nodes to form a communication backplane carrying data between a source and a destination. The data is transported in a multi–hop fashion by a network of nodes that employ untuned narrowband radios. Let $N$ denote the number of unit–capacity channels available for communication. We show that by using random linear network coding, achievable throughput of the network is $\Theta(N)$ as long as the number of hops grows sub–exponentially with $N$, same order as in a fully coordinated network of tuned radios. Moreover, the ratio of the achievable throughput of the untuned network to the achievable throughput of a tuned network is shown to be close to $1/e$. We also show that a scheme in which packets are simply forwarded (without the use of network coding), cannot achieve throughput that grows with $N$, even if the number of hops is only linear in $N$. Note that the same approach can be used over time, with nodes choosing to transmit, receive or sleep in randomly selected timeslots. The precise formulations and proofs of these results can be found in [9], [10], [11], along with a more detailed discussion of this work.

III. DISTRIBUTED NETWORKED STORAGE

The popular approach to retrieving data in wireless sensor networks is for the query unit to ask for the data from the sensor nodes of interest. The desired data is then routed from the source nodes to the data collector. This may be categorized as a "pull-based" strategy that is performed at query time.

The most common approach used to pull the data is to flood queries to all nodes, and construct a spanning tree by having each node maintain a routing table of their parents. This is the approach currently used in both TinyDB and Cougar [8]. In certain scenarios of interest, a pull-based approach at query time may have limitations. Primarily, there can potentially be a large latency in getting the desired data out of a multitude of source nodes scattered randomly across the network due to the multi-hop routing phase following the query. In general, processing done at query time introduces latency and unreliability that may not be acceptable for certain applications.

This work is accordingly motivated at trying to reduce latency and unreliability between query time and the time that the desired data is made available to the data collector. In the spirit of "smart dust" sensor networks, we consider a very large scale network with individual nodes severely constrained by communication, computation, and memory. When one envisions large numbers of cheap, unreliable sensors it is very reasonable to introduce redundancy to ensure that the whole network is acting as a robust distributed storage database. Of particular importance is the assumption that memory cannot scale with the size of the network.

We study the interesting case where a data collection query can be posed anywhere in the network and require quick and reliable access to the data of interest, which is scattered randomly throughout the network. Clearly, this would require some sort of a "pre–routing" phase wherein the network needs to be prepared so as to allow for this low latency query answering capability. The key issue, of course, is whether it is possible to achieve this robust distributed storage under the communication and storage constraints imposed by the nature of wireless sensors.

*A. Main Results*

Formally we define the problem of distributed networked storage (independently introduced in [4]) where there are $k$ data nodes, each producing (e.g. by collecting sensor data readings) one data packet of interest and $n$ storage nodes that will be used as a distributed network memory. We assume that the ratio $k/n$ stays fixed as the network scales and that storage nodes can store only one data packet. We would like to diffuse (i.e. pre–route) the data packets to the $n$ storage nodes so that by querying *any* $k$ storage nodes, it is possible to retrieve all the $k$ data packets of interest (with high probability).

We show [5], [7] how one can create sparse random linear codes to solve this problem by having each data node pre–route its packet to only $O(\log n)$ randomly and independently selected storage nodes. These *decentralized erasure codes* constitute a new class of erasure correcting codes with the key property that they can be created (with minimal communication) when both the initial data and the storage locations are distributed. Mathematically, our problem is equivalent with constructing sparse random graphs which can support sufficiently large flow. We demonstrate that the logarithmic pre–routing degree is optimal if one wants to recover all the $k$ data packets. However if one slightly relaxes the problem and requires to be able to recover $(1-\delta)k$ data packets by querying any $(1+\epsilon)k$ storage nodes (for fixed constants $\epsilon, \delta > 0$), then we demonstrate [6] how one can construct distributed fountain codes [14] to solve this problem with only constant (independent of the network size $n$) pre–routing degrees.